\documentclass[reprint,twoside,twocolumn,nofootinbib]{revtex4}
\usepackage{dcolumn}
\usepackage{latexsym}
\usepackage{amsmath}
\usepackage{amssymb}
\usepackage{amsthm}
\usepackage{dsfont}
\usepackage{longtable}
\usepackage{color}
\usepackage{verbatim}
\usepackage{graphicx}
\usepackage{epstopdf}

\usepackage[pdfstartview={FitH top},pdfpagemode=None,breaklinks]{hyperref}
\hypersetup{
    colorlinks,
    citecolor=black,
    filecolor=black,
    linkcolor=black,
    urlcolor=blue
}

\newcommand {\dd}[2] {\frac {\partial {#1} }{\partial {#2}}}

\newcommand{\Tr}{\mathop{\mathrm{Tr}}\nolimits}

\newcommand{\diag}{\mathop{\mathrm{diag}}\nolimits}

\theoremstyle{definition}

\makeatletter
\@addtoreset{remark}{utv}
\@addtoreset{remark}{theorem}
\@addtoreset{paragraph}{section}
\makeatother

\allowdisplaybreaks
\begin{document}
\title[Energy dependence of the entanglement entropy ...]
{Energy dependence of the entanglement entropy of composite boson (quasiboson) systems}%

\author{A.M. Gavrilik}
\affiliation{Bogolyubov Institute for Theoretical Physics, Nat. Acad. of Sci. of Ukraine}
\address{14b, Metrolohichna Str., Kyiv 03680, Ukraine}
\email{omgavr@bitp.kiev.ua}
\author{Yu.A. Mishchenko}
\affiliation{Bogolyubov Institute for Theoretical Physics, Nat. Acad. of Sci. of Ukraine}
\address{14b, Metrolohichna Str., Kyiv 03680, Ukraine}

\pacs{No. PACS} 

\setcounter{page}{1}%

\begin{abstract}
Bipartite composite boson (quasiboson) systems, which admit
realization in terms of deformed oscillators, were considered in our
previous paper from the viewpoint of entanglement characteristics.
These characteristics, including entanglement entropy and purity,
were expressed through the relevant deformation parameter for
different quasibosonic states. On the other hand, it is of interest
to present the entanglement entropy and likewise the purity as
function of energy for those states. In this work, the corresponding
dependencies are found for different states of composite bosons
realized by deformed oscillators and, for comparison, also for the
hydrogen atom viewed as composite boson. The obtained results are
expressed graphically and their implications discussed.
\end{abstract}
\maketitle

\section{Introduction}

Composite bosons or quasibosons or cobosons, as non-elementary
systems or (quasi-)particles built from two or more constituent
particles, are widely
encountered~\cite{Tichy_Rev,Avan,Hadjimichef,Perkins_2002,Keldysh_Kozlov,Moskalenko2000Bose,Bethe_Salpeter}
in diverse branches of modern theoretical (quantum) physics. Among
quasibosons there are mesons, diquarks/tetraquarks, odd-odd or
even-even nuclei, positronium, excitons, cooperons, atoms, etc. In
the present work we focus on the case of bipartite (two-component)
composite bosons. Their creation and annihilation operators can be
given through the typical ansatz, $A^{\dag}_{\alpha} =
\sum_{\mu\nu}\Phi_{\alpha}^{\mu\nu} a^{\dag}_{\mu}b^{\dag}_{\nu},\ \
A_{\alpha} = \sum_{\mu\nu} \overline{\Phi_{\alpha}^{\mu\nu}} b_{\nu}
a_{\mu}$, where $a^{\dag}_{\mu}$ and $b^{\dag}_{\nu}$ are the
creation operators for the (distinguishable) constituents, which can
be either both fermionic or both bosonic. In~\cite{GKM2,GKM} it was
shown that the composite bosons of particular form (i.e. those that
involve appropriate matrices $\Phi_{\alpha}^{\mu\nu}$) can be
realized, in algebraic sense, by suitable deformed bosons (deformed
oscillators).

As known, among the measures characterizing degree of entanglement
or correlation between the entangled constituents in a quasiboson
there are Schmidt rank, Schmidt number, concurrence, purity, and the
entanglement entropy. The latter two are especially important in the
context of (theoretical and experimental) quantum information
research, quantum communication and
teleportation~\cite{Horodecki,Tichy_Rev}.

It is very important to know how the change of system's energy
influences the quantum correlation and/or quantum statistics
properties of the system under study.
 As it is known, the
characteristics of the entanglement between
 constituents of quasiboson, which measure bosonic quality of
 quasiboson~\cite{Law,Chudzicki,Ramanathan,Morimae}, and their energy dependence
 are of importance in quantum information research: the quantum
communication, entanglement production~\cite{Weder}, quantum
dissociation processes~\cite{Esquivel}, particle addition or
subtraction in general and in teleportation
problem~\cite{Kurzynski,Bartley}, etc. The knowledge of the energy
dependence of witnesses of quantum correlations e.g. entanglement
entropy or purity allows one to relate these latter to the energy
level of excitation that can be measured in the experiments, see
e.g.~\cite{Chang_Photosynth,Navarrete-Benlloch,Esquivel}.

In other words, the energy of a quasiboson differs from the energy
of the respective ideal boson by a term which depends on the
quasiboson's entanglement -- the measure of deviation from bosonic
behavior. All this motivates to study the energy dependence of the
entanglement entropy and other witnesses of entanglement. In the
present work we analyze interconnection between the energy of system
(state) and such main two entanglement characteristics as
entanglement entropy or purity. Note that the relationship between
the entanglement and energy for composite bosons was discussed
in~\cite{Majtey,Bouvrie}, for qubits in~\cite{McHugh,Cavalcanti},
and for spin systems in~\cite{Wang}.

For those composite bosons realizable by deformed oscillators it is
possible, as shown in~\cite{GM_Entang}, to link directly the
relevant parameter of deformation with the entanglement
characteristics of the composite bosons. Namely, the characteristics
(or measures) of bipartite entanglement with respect to $a$- and
$b$-subsystems, see the above ansatz, were found
explicitly~\cite{GM_Entang} for single composite boson, for
multi-quasiboson states, and for a coherent state, corresponding to
the quasibosons system under study.

Among the above mentioned entanglement characteristics the
entanglement entropy $S_{ent}$ certainly is of primary interest.
Therefore in this work main attention is devoted to finding the
explicit dependence of entanglement entropy $S_{ent}$ on the energy
$E$ of the quasibosons system i.e. of the corresponding state.
Present paper further develops the findings of~\cite{GM_Entang}: we
take the composite bosons system as being realized in terms of
independent-modes deformed oscillators with the
quadratic\footnote{As proven in~\cite{GKM,GKM2} this is the only
possibility in case when the both constituents are pure fermions (or
pure bosons)} structure function $\varphi(n)=\Bigl(1+\epsilon
\frac{f}{2}\Bigr)n - \epsilon \frac{f}{2}n^2$, where
$\epsilon=+1/-1$ for fer\-mi\-on\-ic/bos\-on\-ic constituents
respectively. Our analysis here is performed for the states
considered as the examples in~\cite{GM_Entang}, and also for the
hydrogen atom as an independent example.  The obtained dependences
$S_{ent}(E)$ of the entanglement entropy on energy are shown
graphically for a few values of the deformation parameter~$f$; one
of the cases is compared with the situation emerging for the
hydrogen atom.

Analogous treatment, although in a shorter fashion is also performed
for the purity-energy dependence. About the structure of the paper:
in the next Sec.~\ref{sec:prelim} we sketch some facts necessary for
what follows; main results on the energy dependence of the
entanglement entropy of (multi-)quasiboson states are presented in
Sec.~\ref{sec:1quas} and \ref{sec:multi}, whereas similar treatment
for the purity witness of bipartite entanglement is given in
Section~\ref{sec:other_witn}. The paper ends with discussion of the
obtained results and of the most interesting from our viewpoint
physical implications (sec.~\ref{sec:discussion}).

\section{Preliminaries}\label{sec:prelim}

As already mentioned, we deal with composite bosons, which are
realized by mode-independent system of deformed bosons (deformed
oscillators) given for one mode by the structure
function~$\varphi(n)$. That means that {\it algebraically} the
quasiboson operators~$A_{\alpha}$, $A^\dag_{\alpha}$ and the number
operator~$N_\alpha$ satisfy on the states the same relations as the
corresponding deformed oscillator creation, annihilation and
occupation number operators:
\begin{align}
&A^\dag_\alpha A_\alpha = \varphi(N_\alpha),\\
&[A_\alpha,A^\dag_\beta] = \delta_{\alpha\beta} \bigl(\varphi(N_\alpha+1)-
\varphi(N_\alpha)\bigr),\\
&[N_\alpha,A^\dag_\beta] = \delta_{\alpha\beta} A^\dag_\beta,\quad [N_\alpha,A_\beta] =
 - \delta_{\alpha\beta} A_\beta,
\end{align}
where the Kronecker deltas reflect mode-independence. Such a
realization is possible, see~\cite{GKM,GKM2}, only when the
structure function~$\varphi(n)$ involving discrete deformation
parameter $f$ is quadratic in~$n$, namely (recall that
 $\epsilon=\pm 1$)
\begin{equation}\label{phi(n)}
\varphi(n)=\Bigl(1+\epsilon \frac{f}{2}\Bigr)n - \epsilon
\frac{f}{2}n^2,\quad f=\frac2m,\ \ m\in\mathds{N},
\end{equation}
whereas the matrices $\Phi_{\alpha}$, are of the
form:
\begin{equation}
\Phi_{\alpha} = U_1(d_a) \diag\Bigl\{0..0,\sqrt{f/2}\, U_{\alpha}(m),0..0\Bigr\}U^{\dag}_2(d_b).\label{gen_solution}
\end{equation}
Note that the state of one composite boson
\begin{equation}\label{1state}
|\Psi_{\alpha}\rangle \!=\! \sum_{\mu\nu}\Phi_{\alpha}^{\mu\nu}
|a_{\mu}\rangle\!\otimes\!|b_{\nu}\rangle,\ \ |a_{\mu}\rangle\equiv
a^{\dag}_{\mu}|0\rangle,\ |b_{\nu}\rangle\equiv
b^{\dag}_{\nu}|0\rangle,
\end{equation}
see ansatz, is in general bipartite entangled with respect to the
states of two constituent fermions (or two bosons); likewise, the
state describing many composite bosons,
\begin{equation}\label{multi_state}
|\Psi\rangle = \sum\limits_{\{n_{\gamma}\}}
\Psi\bigl(\{n_{\gamma}\}\bigr) (A^{\dag}_{\gamma_1})^{n_{\gamma_1}}
\cdot...\cdot (A^{\dag}_{\gamma_D})^{n_{\gamma_D}}|0\rangle
\end{equation}
is viewed as bipartite entangled with respect to $a$- and
$b$-subsystems. The degree of entanglement can be measured by such
well-known characteristics as Schmidt rank, Schmidt number, purity,
entanglement entropy and concurrence, see
e.g.~\cite{Tichy_Rev,Horodecki} for their definition.

For the entanglement entropy in the case of one composite boson we
obtain~\cite{GM_Entang},
\begin{equation}
S_{\rm ent} = \ln(m) = \ln\frac2f,
\end{equation}
whereas for the multi-quasibosonic states~(\ref{multi_state}),
see~\cite{GM_Entang},
\begin{align}
&S_{\rm ent} = -\sum\limits_{\{n_{\gamma}\}} |\Psi\bigl(\{n_{\gamma}\}\bigr)|^2
\Bigl(\frac1m\Bigr)^{\sum\limits_{j=1}^D n_{\gamma_j}}
\prod\limits_{j=1}^D (n_{\gamma_j}!)^2 N_m^{n_{\gamma_j}} \cdot\nonumber\\
&\cdot\ln \biggl[|\Psi\bigl(\{n_{\gamma}\}\bigr)|^2
\Bigl(\frac1m\Bigr)^{\sum\limits_{j=1}^D n_{\gamma_j}}
\prod\limits_{j=1}^D (n_{\gamma_j}!)^2\biggr].\label{mchar3}
\end{align}

\section{Energy dependence of the entanglement entropy}\label{sec:1quas}

In order to find the energy dependence of the entanglement entropy
we need the expression for the Hamiltonian of the composite boson
system. Different choices are possible here, but, since quasibosons
in our approach are realized by means of deformed oscillators, we
adopt the corresponding Hamiltonian of the same form as e.g.
in~\cite{GR4,GR5}. That is, we take the following Hamiltonian of
deformed oscillators (deformed bosons) which provide realization of
the composite bosons:
\begin{equation}\label{Hamiltonian}
H = \sum_\alpha \frac12 \hbar\omega_\alpha \bigl(\varphi(N_\alpha)+\varphi(N_\alpha+1)\bigr).
\end{equation}

\paragraph{Single composite boson (quasiboson) case.}

As our first example, consider the system which consists of
single composite boson. For the entanglement entropy in this case we
have~\cite{GM_Entang}
\begin{equation}
S_{\rm ent} = \ln\frac2f.
\end{equation}
The expression for the energy of one composite boson as follows
from~(\ref{Hamiltonian}) along with~(\ref{phi(n)}), is
\begin{equation}\label{Energy_1_quasibos}
E = \frac{\hbar\omega}{2} \bigl(\varphi(1)+\varphi(2)\bigr) =
\hbar\omega\Bigl(\frac32-\epsilon\frac{f}{2}\Bigr) = \hbar\omega
\bigl(\frac32-\frac{\varepsilon}{m}\bigr).
\end{equation}
Then for the entanglement entropy characterizing single composite
boson we find
\begin{align}\label{S_1_quasibos}
&S_{\rm ent} = \ln \frac{\epsilon}{\frac32-\frac{E}{\hbar\omega}} =\nonumber\\
&= \left\{
\begin{aligned}
&-\ln\Bigl(\frac32-\frac{E}{\hbar\omega}\Bigr),\ \epsilon=1,\ \ \frac12\le\frac{E}{\hbar\omega}\le\frac32,\\
&-\ln\Bigl(\frac{E}{\hbar\omega}-\frac32\Bigr),\ \epsilon=-1,\ \ \frac32\le\frac{E}{\hbar\omega}\le\frac52.
\end{aligned}
\right.
\end{align}

\begin{figure}[h]
\includegraphics[width=0.8\columnwidth]{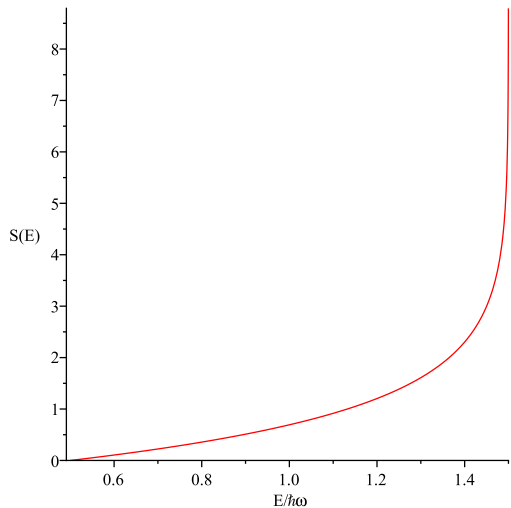}
\vskip-3mm\caption{\footnotesize Dependence of the entanglement
entropy $S_{\rm ent}$ on the energy $E_\alpha$ for a single
composite boson in the case of fermionic components i.e. at
$\epsilon=+1$.} \label{fig1}
\end{figure}

\begin{figure}[h]
\includegraphics[width=0.8\columnwidth]{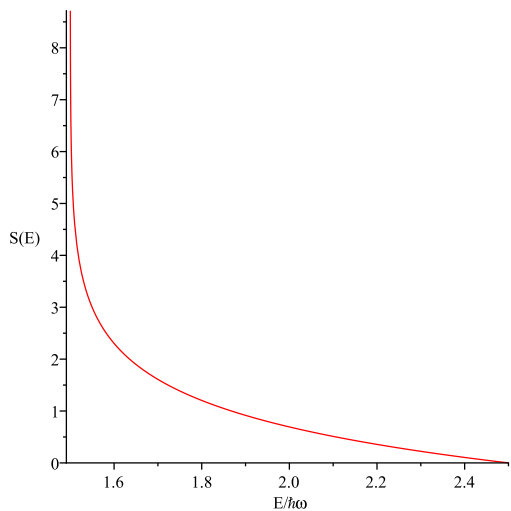}
\vskip-3mm\caption{\footnotesize Dependence of the entanglement
entropy $S_{\rm ent}$ on the energy $E_\alpha$ for a single
composite boson in the case of bosonic components i.e. at
$\epsilon=-1$. } \label{fig2}
\end{figure}

The plots corresponding to eq. (\ref{S_1_quasibos}) are presented
on~Fig.\ref{fig1}, Fig.\ref{fig2}. Note the important feature of the
opposite behavior (increasing vs decreasing) of the energy
dependence in the case of fermionic constituents with respect to the
case of bosonic constituents.
  In the both $\varepsilon=\pm1$ cases the entropy $S_{\rm ent}$ goes to infinity
for the energy $E=\frac32 \hbar\omega$, which implies maximal
entanglement between constituents. In this case the constituents
(fermionic or bosonic) become most tightly bound within a
quasiboson, and the quasiboson is most close to pure boson. On the
contrary for $E=\frac12\hbar\omega$, $\varepsilon=+1$, and
$E=\frac52 \hbar\omega$, $\varepsilon=-1$, the entanglement entropy
$S_{\rm ent}=0$ i.e. the constituents are unentangled.  From
physical viewpoint, in this case the constituents are in fact
unbound.

\paragraph{Hydrogen atom as quasiboson.}

It is of interest to consider the hydrogen atom which constitutes a
composite boson (entangled with respect to proton and electron). In
this case, however, the relevant matrices~$\Phi_{\alpha}^{\mu\nu}$
are not of the form~(\ref{gen_solution}), therefore if it (H-atom)
was realized by a deformed boson (this is an open problem), the
latter should be different from the type mentioned above. So, the
creation operator for the hydrogen atom with zero total momentum and
quantum number~$n$ can be written in second quantization formalism
(with discrete momenta) as\footnote{Note that similar ansatz  is
used for the excitonic creation operators, see
e.g.~\cite{Moskalenko2000Bose,Keldysh_Kozlov}}
\begin{equation}\label{A^dag_Hydr}
A^\dag_{{\bf 0}n} = \frac{(2\pi\hbar)^{3/2}}{\sqrt{V}}  \sum_{\bf p}
\phi_{{\bf p}n} a^\dag_{\bf p} b^\dag_{-\bf p},
\end{equation}
where $a^\dag_{\bf p}$ and $b^\dag_{-\bf p}$ are the creation
operators for electron and proton respectively taken with opposite
momenta; $V$ is large enough confining volume for the hydrogen atom.
The momentum-space wavefunction $\phi_{{\bf p}n}$ is determined by
the Schrodinger equation:
\begin{align*}
&\phi_{{\bf p}n} = \int \frac{1}{(2\pi\hbar)^{3/2}} e^{\frac{i}{\hbar}{\bf pr}}
 \phi_n({\bf r}) d^3{\bf r};\\
&-\frac{\hbar^2\nabla^2}{2m}\phi_n({\bf r}) + U({\bf r})
\phi_n({\bf r}) = E_n \phi_n({\bf r}).
\end{align*}
The expression for the Hydrogen wavefunction in the momentum
representation is given as~\cite{Bethe_Salpeter}
\begin{align}
&\phi_{{\bf p}nlm} \!=\! \frac{e^{\pm im\phi_p}}{(2\pi)^{1/2}}
\Bigl(\frac{(2l\!+\!1)(l\!-\!m)!}{2(l\!+\!m)!}\Bigr)^{1/2}
P^m_l(\cos \theta_p) \frac{\pi 2^{2l+4}l!}{(\gamma h)^{3/2}}\cdot\nonumber\\
&\cdot\Bigl(\frac{n(n\!-\!l\!-\!1)!}{(n\!+\!l)!}\Bigr)^{1/2}
\frac{\xi^l}{(\xi^2\!+\!1)^{l+2}}
C^{l+1}_{n-l-1}\Bigl(\frac{\xi^2\!-\!1}{\xi^2\!+\!1}\Bigr),\label{phi_nlm}
\end{align}
where $P^m_l$ is the associated Legendre polynomial, $C^{l+1}_{n-l-1}(...)$
is Gegenbauer polynomial, $\xi = (2\pi/\gamma h)p$, $\gamma=Z/na_0$.

The expansion~(\ref{A^dag_Hydr}) can be viewed directly as the
Schmidt decomposition for the state~$A^\dag_{{\bf 0}n}|0\rangle$
with Schmidt coefficients~$\lambda_{\bf p} =
\frac{(2\pi\hbar)^{3/2}}{\sqrt{V}} \phi_{{\bf p}n}$. Then the
entanglement entropy for the hydrogen atom is given by the relation
\begin{align}
&S_{\rm ent} =  -\sum_{\bf p} |\lambda_{\bf p}|^2 \ln |\lambda_{\bf
p}|^2 =\nonumber\\
&= - \sum_{\bf p} \frac{(2\pi\hbar)^3}{V} |\phi_{{\bf p}n}|^2 \ln
\Bigl(\frac{(2\pi\hbar)^3}{V} |\phi_{{\bf
p}n}|^2\Bigr),\label{S_Hydr}
\end{align}
where the first equality is nothing but the definition of the
entanglement entropy~\cite{Tichy_Rev}.

\begin{figure}[h]
\includegraphics[width = 0.8\columnwidth]{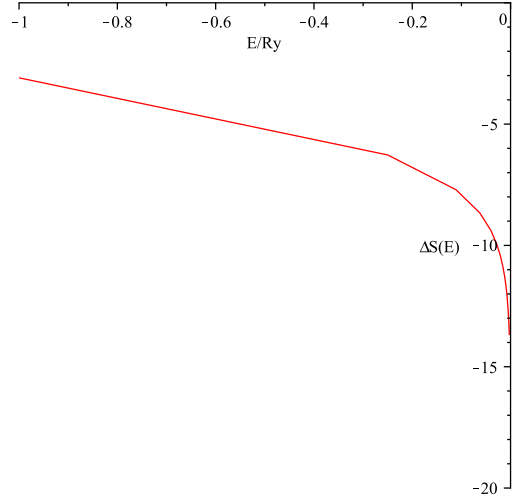}
\vskip-3mm\caption{\footnotesize Dependence of the entanglement
entropy $\Delta S = S_{\rm ent}-S_{\rm ent}^{(0)}$
from~~(\ref{S_ent4}) on the energy $E$ for Hydrogen atom.}
\label{fig3}
\end{figure}

Calculation of the expression~(\ref{S_Hydr}) is moved to appendix.
Performing derivation we obtain the following result
\begin{multline}
S_{\rm ent} = - \ln\Bigl[\frac{(2l+1)(l-m)!}{(l+m)!}\frac{4\pi 2^{2l}(l!)^2}{V(na_0)^{-3}} \frac{n(n-l-1)!}{(n+l)!}\Bigr]-\\
- \frac{(2l+1)(l-m)!}{2(l+m)!} \, \int\limits_{-1}^{1} dt |P^m_l(t)|^2 \ln |P^m_l(t)|^2 -\\
\!-\!\frac{4^l(l!)^2}{\pi/2} \frac{n(n\!-\!l\!-\!1)!}{(n+l)!}
\!\int\limits_{-1}^{1}\! dx \frac{\sqrt{1\!-\!x^2}\!}{(1\!-\!x)^3}
G_{nl}\!(x) \ln G_{nl}\!(x),
\end{multline}
where $G_{nl}(x) = (1-x^2)^l (1-x)^4 \bigl(C^{l+1}_{n-l-1}(x)\bigr)^2$.
Let us consider the simplest case when the quantum numbers $l=0$ and $m=0$. For these values,
\begin{multline}\label{S_ent2}
S_{\rm ent} = \ln\Bigl[\frac{V}{4\pi n^3a_0^3}\Bigr]
- \frac{2}{\pi} \int\limits_{-1}^{1} dx (1-x^2)^{1/2} (1-x) \bigl(C^{1}_{n-1}(x)\bigr)^2\cdot\\
\cdot \ln\Bigl[(1-x)^4 \bigl(C^{1}_{n-1}(x)\bigr)^2\Bigr]
= S_{\rm ent}^{(0)} - \ln[4\pi n^3] -\\
- \frac{2}{\pi} \!\int\limits_{-1}^{1}\! dx(1-x^2)^{1/2} (1-x) \bigl(C^{1}_{n\!-\!1}(x)\bigr)^2\cdot\\
\cdot\ln\Bigl[(1-x)^4 \bigl(C^{1}_{n\!-\!1}(x)\bigr)^2\Bigr],\qquad
S_{\rm ent}^{(0)} = \ln\frac{V}{a_0^3}.
\end{multline}
Making replacement $x=\cos\alpha$ in the integral in~(\ref{S_ent2}),
and using the formula $C_{n-1}^1(\cos\alpha) =
\frac{\sin(n\alpha)}{\sin\alpha}$, we infer:
\begin{multline}\label{S_ent3}
S_{\rm ent} = S_{\rm ent}^{(0)} - \ln[4\pi n^3]
- \frac{2}{\pi} \int\limits_{0}^{\pi} d\alpha (1-\cos\alpha) \sin^2(n\alpha)\cdot\\
\cdot \ln\Bigl[(1-\cos\alpha)^4
\frac{\sin^2(n\alpha)}{\sin^2\alpha}\Bigr]
\end{multline}
From the well-known expression for the energy of H-atom, $E=-{\rm
Ry}/n^2$, we have $n = \sqrt{-{\rm Ry}/E}$.
 Substituting this in~(\ref{S_ent3}) we finally obtain
\begin{multline}\label{S_ent4}
\Delta S(E) = S_{\rm ent} (E) - S_{\rm ent}^{(0)}
= - \ln\Bigl[4\pi \Bigl(-\frac{\rm Ry}{E}\Bigr)^{3/2}\Bigr]-\\
- \frac{2}{\pi} \int\limits_{0}^{\pi} d\alpha (1-\cos\alpha)
 \sin^2\Bigl(\sqrt{-\frac{\rm Ry}{E}}\alpha\Bigr)\cdot\\
\cdot\ln\Bigl[(1-\cos\alpha)^4 \frac{\sin^2\bigl(\sqrt{-\frac{\rm
Ry}{E}}\alpha\bigr)}{\sin^2\alpha}\Bigr]
\end{multline}
The derived energy dependence is shown graphically on
Fig.~\ref{fig3}. As seen, the character of the energy dependence
here essentially differs from that of the single quasiboson
(two-fermion composite) case above, see~Fig.\ref{fig1}. Main reason
for distinction lies in that the matrices $\Phi_\alpha^{\mu\nu}$ of
composite bosons realized by deformed oscillators are different from
the corresponding matrices of hydrogen atom given by its
wavefunction~$\phi_{{\bf p}n}$. From the physics viewpoint this
implies that the effective interaction between the constituents in
the above quasiboson is different from the Coulomb interaction
within hydrogen atom.

Of course, it would be useful to perform the analysis of H-atom
system by taking into account the fact that the proton, in its turn,
also has composite structure (three-quark system), thus differing
from fundamental (elementary) fermionic entity.

Let us recall once more that the example of H-atom, treated as
quasiboson, is included here for comparative purpose only. We
suppose however that some realization (by deformed oscillators) for
hydrogen atom as composite quasiboson does also exist, though it may
be rather complicated and non-algebraic. In fact, this task is to be
solved in two stages: first we have to construct a realization of
3-fermion (3-quark) state for the proton subsystem by some deformed
fermionic oscillator, and, second, of a quasi-boson formed from the
(protonic) quasifermion and one more fermion, i.e. electron. We hope
to solve this very involved problem in near future.

\section{Entanglement entropy vs energy for multi-quasiboson
system}\label{sec:multi}

Now examine the case of multi-quasiboson states. Taking into account
the Hamiltonian~(\ref{Hamiltonian}), the total energy of the system
(at mode-independence) is expressed as
\begin{equation}\label{E_n}
E = \sum_\alpha \hbar\omega_\alpha\Bigl[n_\alpha + \frac12 -
\epsilon\frac{f}{2} n_\alpha^2\Bigr].
\end{equation}

\paragraph{Qusiboson Fock state.}

\begin{figure}[h!]
\includegraphics[width = 0.8\columnwidth]{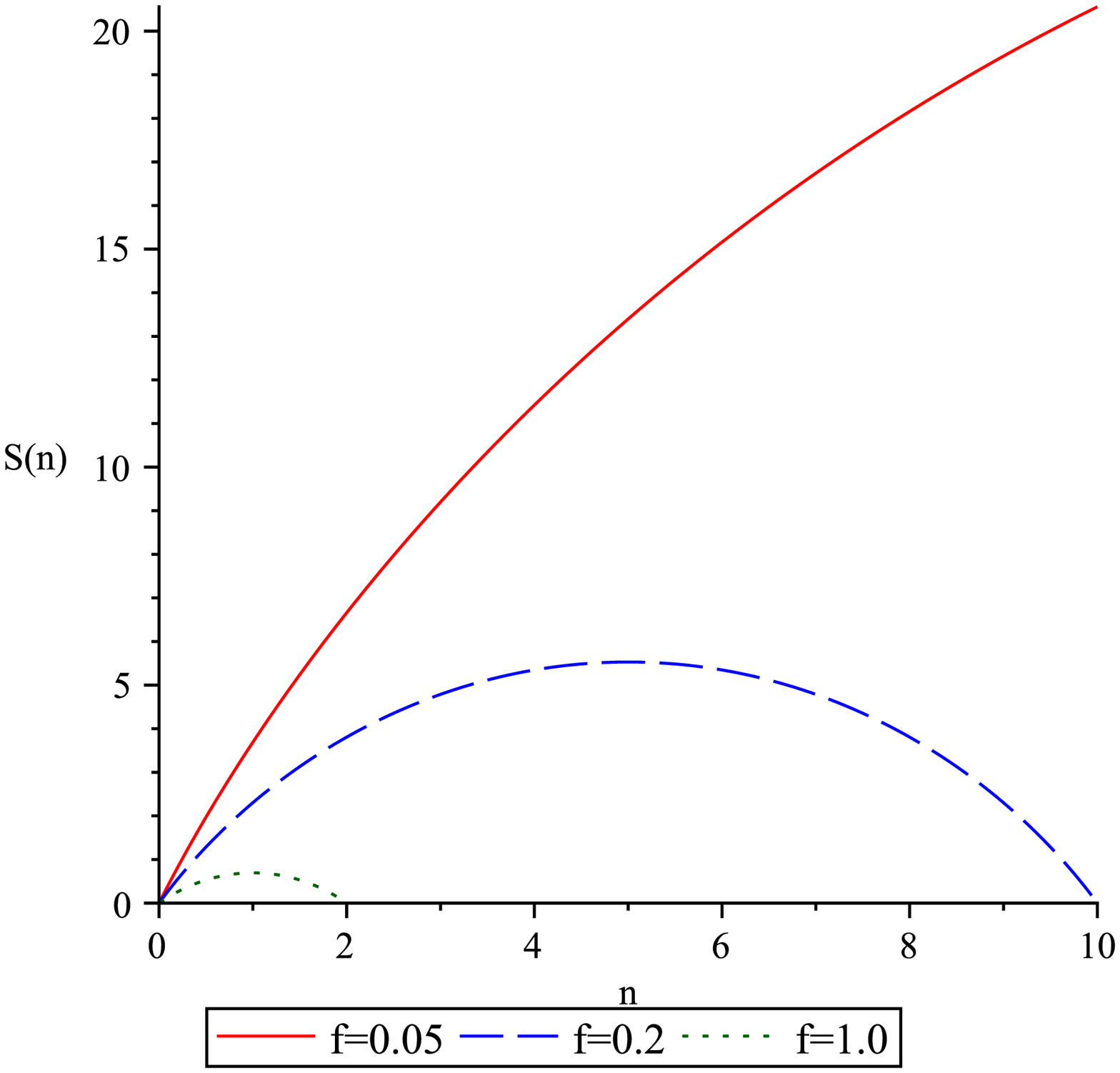}
\vskip-3mm\caption{\footnotesize Dependence of the entanglement
entropy $S_{\rm ent}$, see~(\ref{S_ex1}), on the number of
quasibosons $n_\alpha$ for one-mode multi-quasibosonic system: the
case $\epsilon=+1$ of fermionic components.} \label{fig4}
\end{figure}

\begin{figure}[h!]
\includegraphics[width = 0.8\columnwidth]{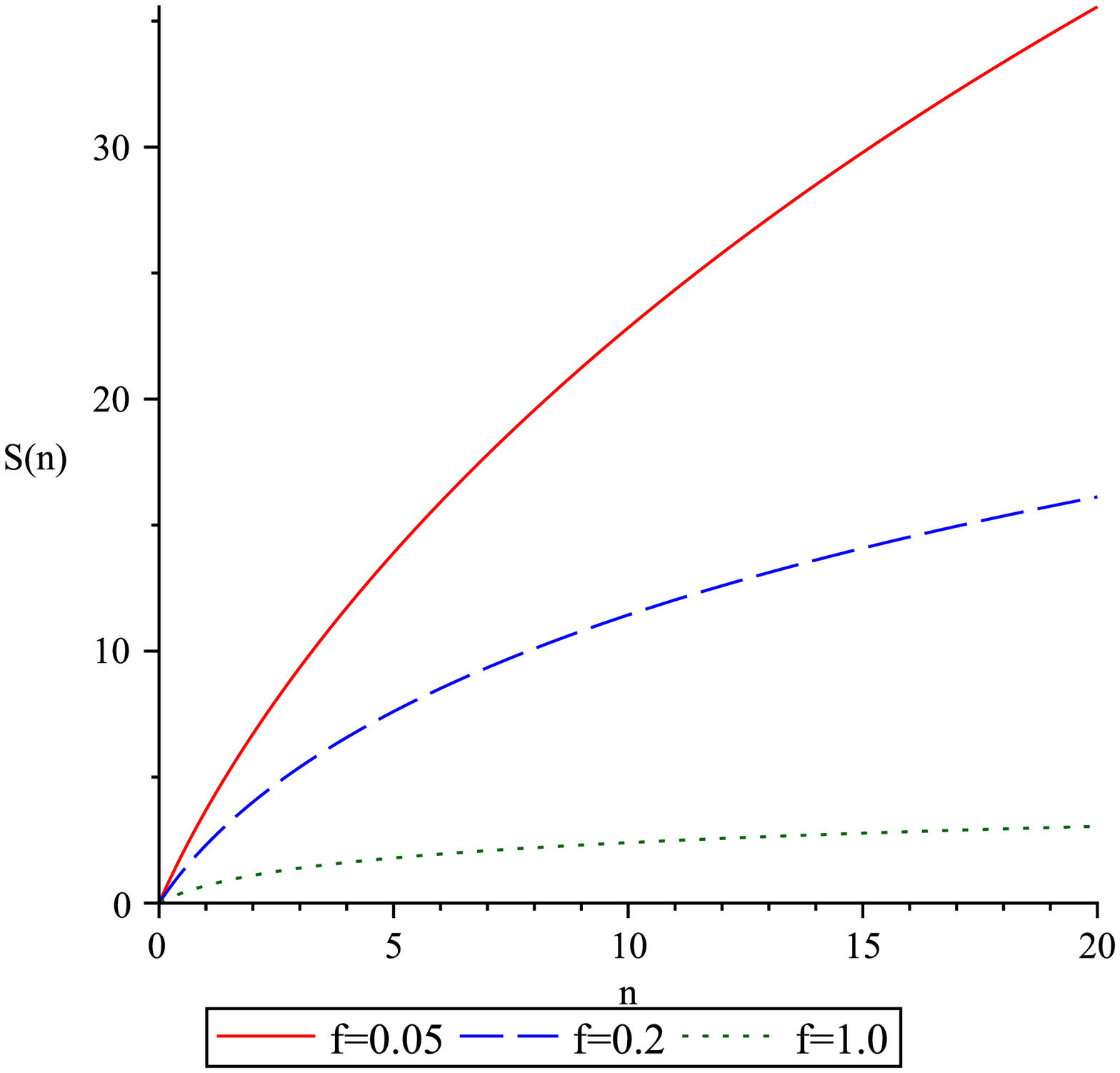}
\vskip-3mm\caption{\footnotesize Dependence of the entanglement
entropy $S_{\rm ent}$, see~(\ref{S_ex1}), on the number of
quasibosons $n_\alpha$ for one-mode multi-quasibosonic system: the
case $\epsilon=-1$ of bosonic components. } \label{fig5}
\end{figure}

Let us find the entanglement entropy as function of energy for the
normalized Fock state of $n_\alpha$ quasibosons,
$[\phi(n_{\alpha})!]^{-1/2}(A^{\dag}_{\alpha})^{n_{\alpha}}|0\rangle$,
in a fixed mode $\alpha$. The entropy of entanglement between $a$-
and $b$-subsystems for the two values of $\epsilon$ equals
respectively, see~\cite{GM_Entang}, to
\begin{equation}\label{S_ex1}
S_{\rm ent}|_{\epsilon=+1}= \ln C_{2/f}^{n_{\alpha}} ,\quad   S_{\rm
ent}|_{\epsilon=-1}= \ln C_{2/f+n_{\alpha}-1}^{n_{\alpha}}.
\end{equation}
The latter dependencies $S_{\rm ent} = S_{\rm ent}(n_\alpha)$ are
shown in Fig.~\ref{fig4} and Fig~\ref{fig5}.

\begin{figure}[h!]
\includegraphics[width = 0.8\columnwidth]{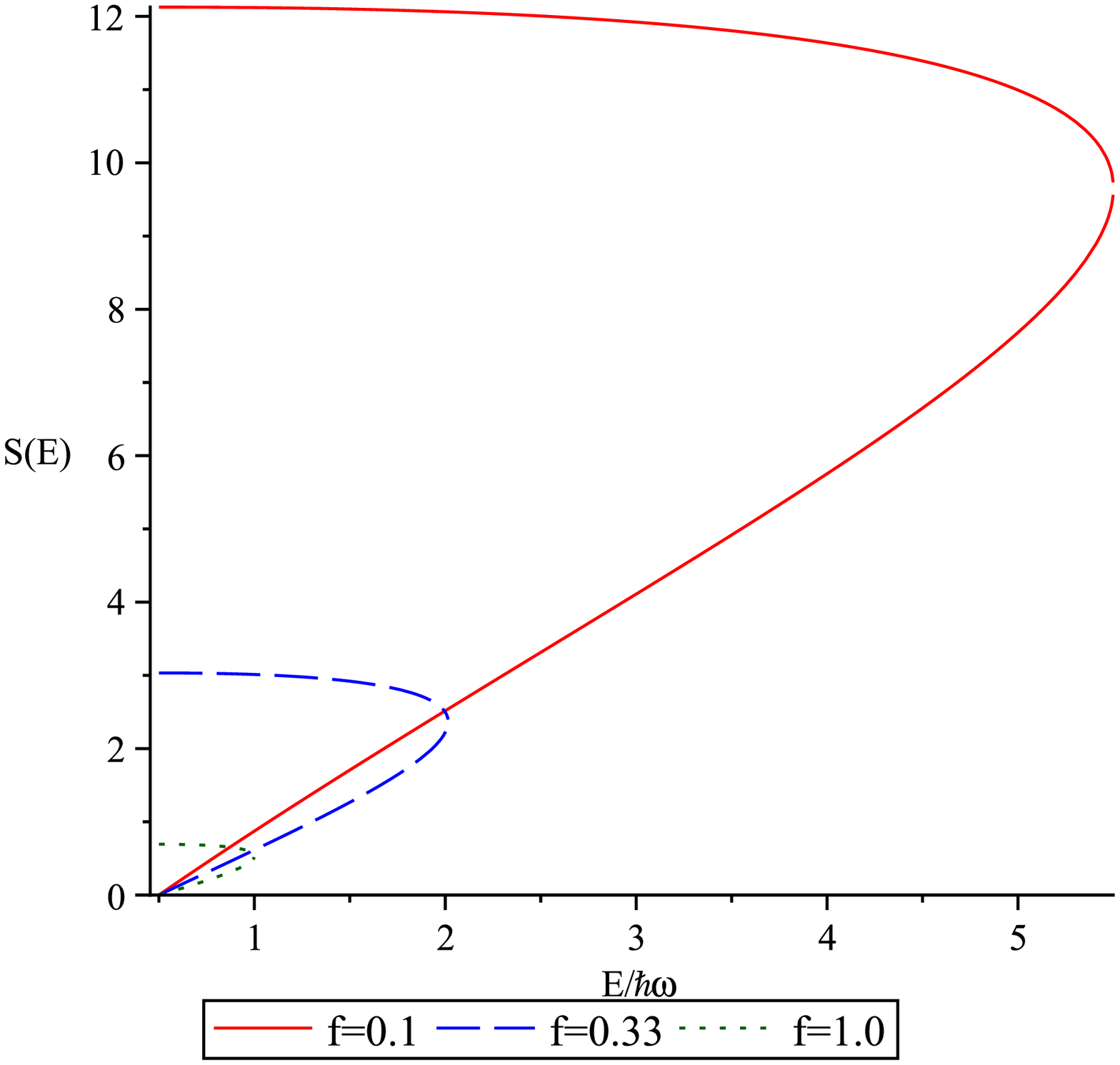}
\vskip-3mm\caption{\footnotesize Dependence of the entanglement
entropy $S_{\rm ent}$, see~(\ref{S_1quas_ferm}), on the energy
$E_\alpha$ for one-mode multi-quasibosonic system: the case
$\epsilon=+1$ of fermionic components.} \label{fig6}
\end{figure}

\begin{figure}[h!]
\includegraphics[width = 0.8\columnwidth]{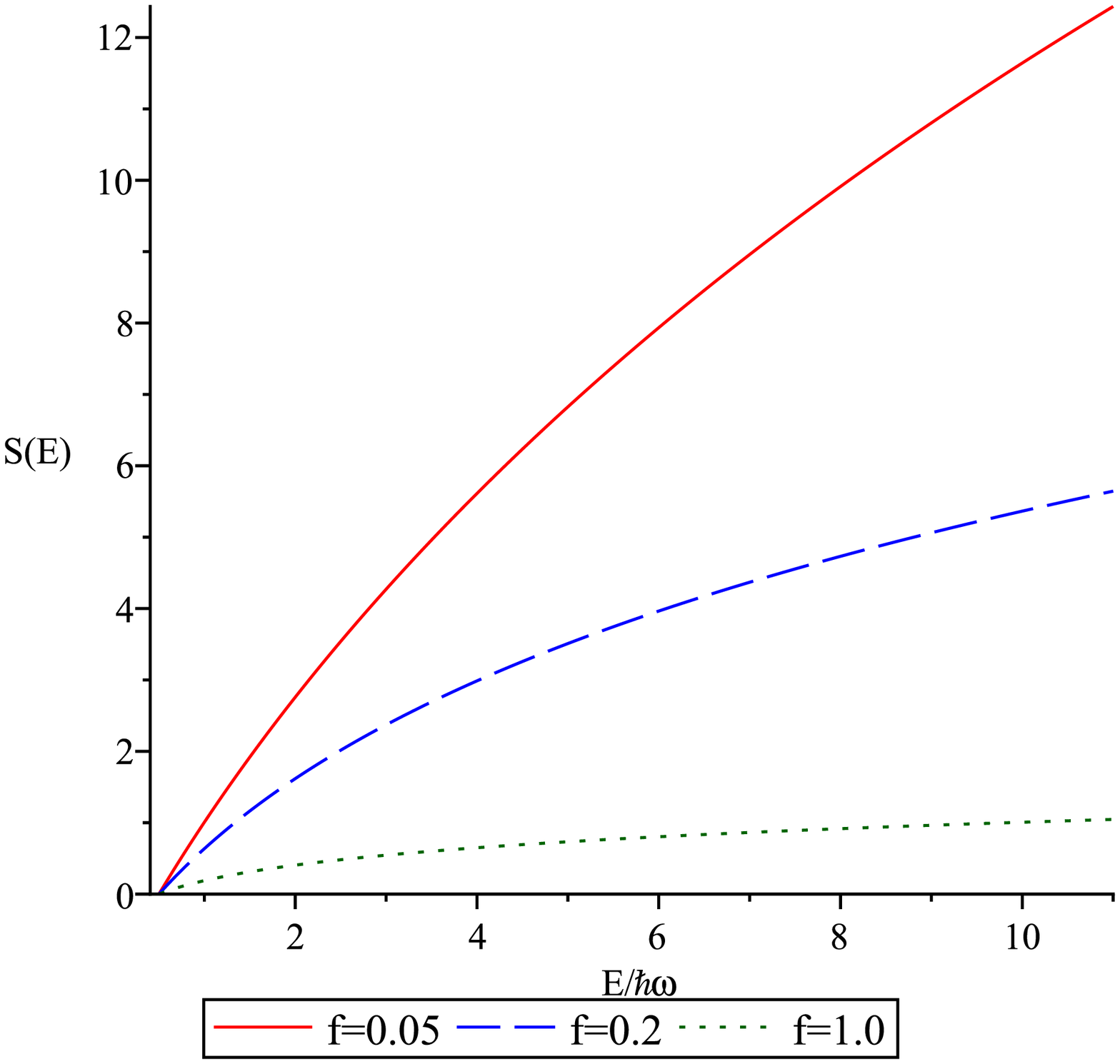}
\vskip-3mm\caption{\footnotesize Dependence of the entanglement
entropy $S_{\rm ent}$, see~(\ref{S_1quas_bos}), on the energy
$E_\alpha$ for one-mode multi-quasibosonic system: the case
$\epsilon=-1$ of bosonic components. } \label{fig7}
\end{figure}

By inverting eq.~(\ref{E_n}) we have the dependence of the
occupation number $n_\alpha$ of quasibosons in $\alpha$th mode on
the corresponding energy $E_\alpha$ of quasibosons:
\begin{equation*}
n^{\pm}_\alpha(E_\alpha) =  \frac{1\pm \sqrt{1-2\epsilon
f\bigl(\frac{E_\alpha}{\hbar\omega_\alpha}-\frac12\bigr)}}{2\epsilon
f}.
\end{equation*}
Substitution of this expression in~(\ref{S_ex1}) leads us to the
two-branch form of the concerned dependence $S^{\pm}_{\rm
ent}(E_\alpha)$ for the case $\epsilon=+1$:
\begin{equation}\label{S_1quas_ferm}
S^\pm_{\rm ent}(E_\alpha)|_{\epsilon=+1} = \ln \Bigl(C_{2/f}^{[1\pm
\sqrt{1\!+\!f\!-\!2f E_\alpha/\hbar\omega_\alpha}\,]/(2f)}\Bigr),
\end{equation}
where $\frac{E}{\hbar\omega}\le \frac{1\!+\!f}{2f}$ for both
$S^+_{\rm ent}$ and $S^-_{\rm ent}$ branches, and
$\frac{E}{\hbar\omega}\ge\frac12$ for the $S^-_{\rm ent}$-branch.
For $\epsilon=-1$ we have single monotonous branch:
\begin{equation}\label{S_1quas_bos}
S_{\rm ent}(E_\alpha)|_{\epsilon=-1} \!=\! \ln
\Bigl(\!C_{2/f\!-\!1+[\sqrt{1\!-\!f\!+\!2f
E_\alpha/\hbar\omega_\alpha}\!-\!1]/(2f)}^{[\sqrt{1\!-\!f\!+\!2f
E_\alpha/\hbar\omega_\alpha}-1]/(2f)}\!\Bigr),
\end{equation}
where $\frac{E}{\hbar\omega}\!\ge\!\frac12$.
  The corresponding functions are presented graphically in~Fig.~\ref{fig6} and
Fig.~\ref{fig7}.


\paragraph{The state with one quasiboson per mode.}

Now let us turn to the Example~2 from~\cite{GM_Entang}. In this case the quasibosons are all in different modes, i.e. the quasibosonic system is in the state
\begin{equation*}
|\Psi\rangle = A^{\dag}_{\gamma_1}\cdot...\cdot
A^{\dag}_{\gamma_n}|0\rangle,\quad\gamma_i\neq \gamma_j,\ i\neq j,\
i,j=1,...,n.
\end{equation*}
For the entanglement entropy, for $\epsilon=\pm1$, we have
\begin{equation}\label{S_ex2}
S_{\rm ent} = n\ln (m) = n \ln\frac{2}{f}.
\end{equation}
The energy of the system depends on the dispersion relation of
$\omega_{\gamma_j}$ as function of $\gamma_j$. Taking it in linear (in~$\gamma_j$) form, namely $\omega_{\gamma_j} =
\omega_0+(\gamma_j-\gamma_1)\dd{\omega}{\gamma}$, and also
using~(\ref{E_n}) and $n_{\gamma_j}=1$, we arrive at the following
expression for the energy:
\begin{equation}
E = \frac{3-\epsilon f}{2} \bigl(\hbar\omega_0 n + \frac12
\hbar\omega n(n-1)\bigr),
\end{equation}
where $\delta\omega=\dd{\omega}{\gamma}\delta\gamma$.
 Solving the latter yields $n$ as function of energy, namely
\begin{equation}
n(E) = \frac{-1+\frac12 \frac{\delta\omega}{\omega_0} +
\sqrt{\bigl(1-\frac12 \frac{\delta\omega}{\omega_0}\bigr)^2 +
4\frac{\delta\omega}{\omega_0} \frac{1}{3-\epsilon\! f}
\frac{E}{\hbar\omega_0}}}{\delta\omega/\omega_0} .
\end{equation}
Then (\ref{S_ex2}) yields
\begin{equation}
S_{\rm ent}(E) \!=\! \frac{-1\!+\!\frac12
\frac{\delta\omega}{\omega_0} \!+\! \sqrt{\bigl(1\!-\!\frac12
\frac{\delta\omega}{\omega_0}\bigr)^2 \!+\!
4\frac{\delta\omega}{\omega_0} \frac{1}{3-\epsilon\! f}
\frac{E}{\hbar\omega_0}}}{\delta\omega/\omega_0} \,\ln\frac{2}{f}.
\end{equation}

\begin{figure}[h!]
\includegraphics[width = 0.8\columnwidth]{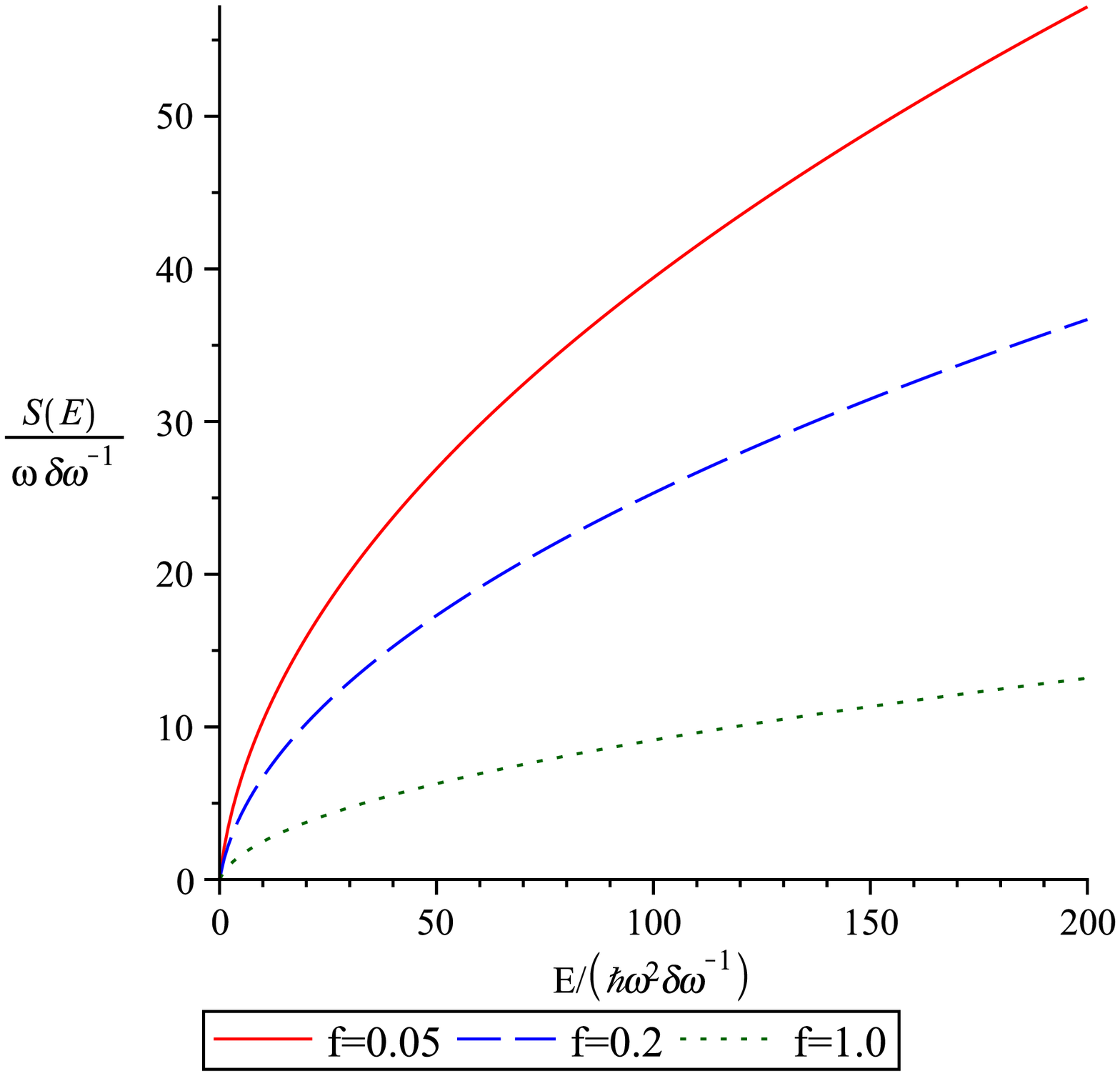}
\vskip-3mm\caption{\footnotesize Dependence of the entanglement
entropy $S_{\rm ent}$ on energy $E$ for multi-quasibosonic system
with one quasiboson per mode: the case $\epsilon=+1$ of fermionic
constituents.} \label{fig8}
\end{figure}

\begin{figure}[h!]
\includegraphics[width = 0.8\columnwidth]{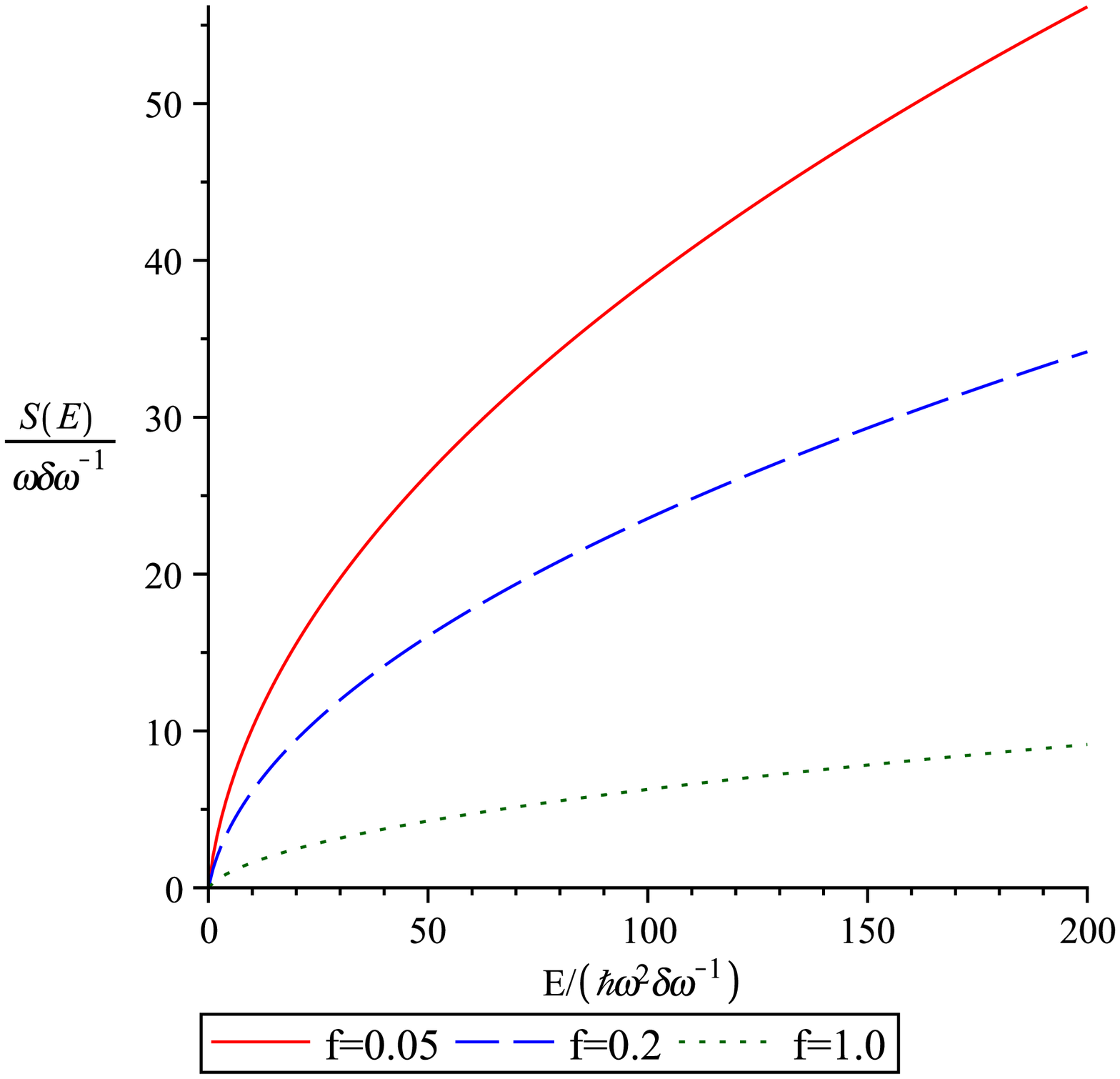}
\vskip-3mm\caption{\footnotesize Dependence of the entanglement
entropy $S_{\rm ent}$ on energy $E_\alpha$ for multi-quasibosonic
system with one quasiboson per mode: the case $\epsilon=-1$ of
bosonic constituents. } \label{fig9}
\end{figure}

Like in the previous case we obtain the corresponding plots which
are now placed in Fig.~\ref{fig8} and Fig.~\ref{fig9} ($S$ is given
in units of $\omega/\delta\omega$ and $E$ in units of
$\hbar\omega^2/\delta\omega$; besides, $\omega
 = |\omega_0 - \frac12 \delta\omega|$ is put).

).

\paragraph{Coherent state of quasibosons.}

\begin{figure}[h!]
\includegraphics[width = 0.8\columnwidth]{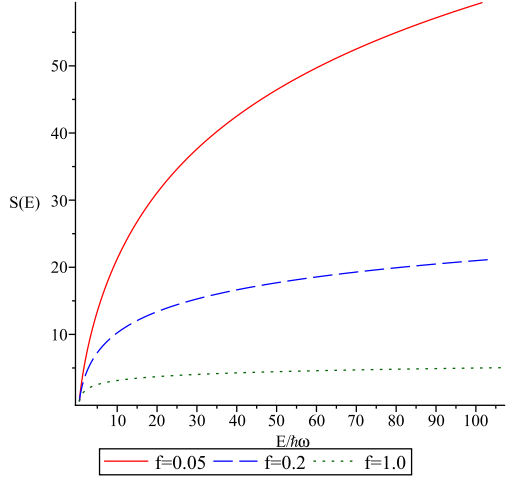}
\vskip-3mm\caption{\footnotesize Dependence of the entanglement
entropy $S_{\rm ent}$ in (\ref{S_coherent}) on the mean energy
$E_\alpha$, see (\ref{E_coherent}), for quasibosonic coherent
state.} \label{fig10}
\end{figure}

As our last example consider the coherent state of composite bosons
system in $\alpha$th mode, see Example~3 in~\cite{GM_Entang}:
\begin{align}
&|\Psi_{\alpha}\rangle \!=\! \tilde{C}(\mathcal{A};m)
 \sum\limits_{n=0}^\infty \frac{\mathcal{A}^n}{\phi(n)!}
 (A^{\dag}_{\alpha})^n|0\rangle,\label{coherent_state}\\
&\tilde{C}(\mathcal{A};m) \!=\! \biggl(\sum\limits_{n=0}^\infty
\frac{|\mathcal{A}|^{2n}}{\phi(n)!}\biggr)^{-1/2} \!\!=\!
\biggl[\frac{(m\!-\!1)! I_{m-1}(z)}{(z/2)^{m-1}}\biggr]^{-\frac12}=\nonumber\\
&= \mathrm{e}^{-|\mathcal{A}|^2/2}\Bigl[1+\frac14
\frac{|\mathcal{A}|^4}{m}+...\Bigr],\ \ \ z=2\sqrt{m}|\mathcal{A}|,\nonumber
\end{align}
where $I_{m-1}(z)$ is the modified Bessel function of order
$m\!-\!1$. For mean energy of the system in this state we have
\begin{multline}\label{E_coherent}
 E_\alpha = \langle\Psi_\alpha| \frac12\hbar\omega_\alpha
 [\varphi(N_\alpha)+\varphi(N_\alpha+1)]|\Psi_\alpha\rangle =
 \frac12 \hbar\omega_\alpha |\tilde{C}|^2\cdot\\
\cdot\sum_{n=0}^{\infty} \frac{|\mathcal{A}|^{2n}}{\varphi(n)!}
\varphi(n) + \frac12\hbar\omega_\alpha |\tilde{C}|^2 \sum_{n=0}^{\infty}
\frac{|\mathcal{A}|^{2n}}{\varphi(n)!} \varphi(n+1) =\\
= \hbar\omega_\alpha|\tilde{C}|^2 |\mathcal{A}|^2 \sum_{n=0}^{\infty}
\frac{|\mathcal{A}|^{2n}}{\varphi(n)!} + \frac12\hbar\omega_\alpha |\tilde{C}|^2
\sum_{n=0}^{\infty} \frac{|\mathcal{A}|^{2n}}{\varphi(n)!}\cdot\\
\cdot [\varphi(n+1)-\varphi(n)] = \hbar\omega_\alpha |\mathcal{A}|^2 +
\frac12\hbar\omega_\alpha |\tilde{C}|^2 \sum_{n=0}^{\infty}
\frac{|\mathcal{A}|^{2n}}{\varphi(n)!}\Bigl[1+\frac{2n}{m}\Bigr]\\
= \hbar\omega_\alpha |\mathcal{A}|^2 + \frac12\hbar\omega_\alpha +
\frac12\hbar\omega_\alpha\frac1m |\tilde{C}|^2 |\mathcal{A}| \dd{}{|\mathcal{A}|}
\sum_{n=0}^{\infty} \frac{|\mathcal{A}|^{2n}}{\varphi(n)!} =\\
= \hbar\omega_\alpha \bigl(|\mathcal{A}|^2+1/2\bigr) +
\frac12 \hbar\omega_\alpha\frac1m |\mathcal{A}|\dd{}{|\mathcal{A}|}
\ln \frac{I_{m-1}(z)}{|\mathcal{A}|^{m-1}} =\\
= \hbar\omega_\alpha\bigl(|\mathcal{A}|^2+\frac12\bigr) +
\hbar\omega_\alpha\frac{1}{\sqrt{m}} |\mathcal{A}| \frac{I'_{m-1}(z)}{I_{m-1}(z)} -
\frac12\hbar\omega_\alpha\frac1m(m-1) =\\
= \hbar\omega_\alpha\Bigl(|\mathcal{A}|^2\!+\!\frac{1}{2m}\Bigr) +
\frac{\hbar\omega_\alpha |\mathcal{A}|}{\sqrt{m}}
\frac{I_m(2\sqrt{m}|\mathcal{A}|)\!+\!I_{m-2}(2\sqrt{m}|\mathcal{A}|)}{2I_{m-1}
(2\sqrt{m}|\mathcal{A}|)}
\end{multline}
The entanglement entropy for the coherent
state~(\ref{coherent_state}) is given, see~\cite{GM_Entang}, by the
 formula (recall that $m=\frac2f$):
\begin{equation}\label{S_coherent}
S_{\rm ent} =\tilde{C}^2 \sum\limits_{n=0}^\infty
\frac{(|\mathcal{A}|^2m)^n}{(n!)^2  C^n_{n+m-1}} \ln
\biggl[\frac{(n!)^2(C_{n+m-1}^n)^2}{\tilde{C}^2
(|\mathcal{A}|^2m)^n}\biggr].
\end{equation}
Hence we have nothing but the dependence of $S_{\rm ent}$ on
$E_\alpha$ in parametric form (unfortunately, we cannot
solve~(\ref{E_coherent}) for $|\mathcal{A}|$, here $|\mathcal{A}|$
being the parameter, in order to insert the solution
into~(\ref{S_coherent}); that is why we merely use parametric
presentation of the~$S_{\rm ent}=S_{\rm ent}(E)$ dependence). The
plot of this dependence is given in Fig.~\ref{fig10}.

\section{Energy dependence of other measures (witnesses) of entanglement} \label{sec:other_witn}

There exist some other widely used witnesses of entanglement:
Schmidt rank, concurrence, Schmidt number $K$ or its inverse $P=1/K$
termed {\it purity} \cite{Tichy_Rev,Horodecki}. Energy dependence of
these entanglement witnesses, Schmidt rank, concurrence and purity
have somewhat simpler form and can be calculated in a similar way
using explicit formulas from~\cite{GM_Entang}.

Since such entanglement characteristics as {\it purity} is exploited
in connection with the issue of entanglement creation in
 scattering processes~\cite{Weder} and others~\cite{Kurzynski,McHugh},
 let us pay some attention to $P$.

For the {\it entangled system consisting of one quasiboson} the
purity in~\cite{GM_Entang} was found to be connected with the
deformation parameter $m=\frac2f$ in a simple way:
\begin{equation}\label{pur_def}
P\!=\!\sum_k \lambda_k^4\!=\!\frac1m,\ \text{or}\ P\!=\!\Tr
(\rho_{\alpha}^{(a)})^2 \!=\! \Tr (\rho_{\alpha}^{(b)})^2 \!=\!
\frac1m.
\end{equation}

Then, the energy dependence for purity in case of a single composite
boson readily follows by combining~(\ref{pur_def}) with
(\ref{Energy_1_quasibos}) that gives:
\begin{align}
&P = \frac{f}{2} = \frac{1}{\epsilon}\Bigl(\frac32-\frac{E}{\hbar\omega}\Bigr) =\nonumber\\
&= \left\{
\begin{aligned}
&\Bigl(\frac32-\frac{E}{\hbar\omega}\Bigr),\ \epsilon=1,\ \
\frac12\le\frac{E}{\hbar\omega}\le\frac32,\\
&\Bigl(\frac{E}{\hbar\omega}-\frac32\Bigr),\ \epsilon=-1,\ \
\frac32\le\frac{E}{\hbar\omega}\le\frac52.
\end{aligned}
\right.
\end{align}
Thus, the dependence of purity on energy \underline{is linear} for
both $\varepsilon=+1$ and $\varepsilon=-1$.
Observe however the two mutually opposite (i.e. falling versus
raising) types of behavior of purity with increasing energy in the
cases of fermionic versus bosonic constituents.

\begin{figure}[h!]
\includegraphics[width = 0.8\columnwidth]{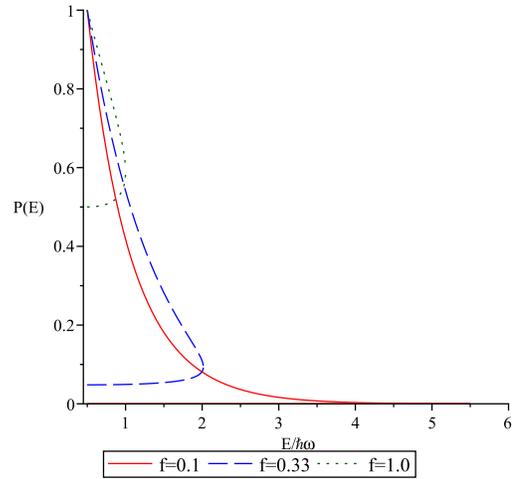}
\vskip-3mm\caption{\footnotesize Dependence of the purity $P$, see
(\ref{P_Fock_ferm}), on the energy $E_\alpha$ for one-mode
multi-quasibosonic system: the case $\epsilon=+1$ of fermionic
components.} \label{fig11}
\end{figure}

\begin{figure}[h!]
\includegraphics[width = 0.8\columnwidth]{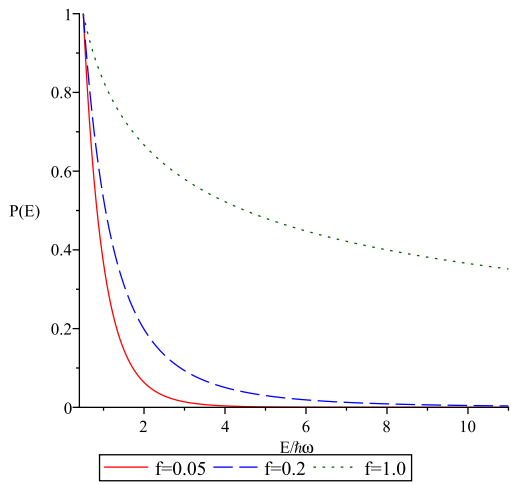}
\vskip-3mm\caption{\footnotesize Dependence of the Purity $P$, see
(\ref{P_Fock_bos}), on the energy $E_\alpha$ for one-mode
multi-quasibosonic system: the case $\epsilon=-1$ of bosonic
components. } \label{fig12}
\end{figure}

In a similar way, for purity in {\it the case of single-mode
multi-quasibosonic Fock states} on the base of~\cite{GM_Entang} we
obtain
\begin{align}
&P^\pm_{\rm ent}(E_\alpha)|_{\epsilon=+1}
= \Bigl(C_{2/f}^{[1\pm \sqrt{1\!+\!f\!-\!2f E_\alpha/\hbar\omega_\alpha}\,]/(2f)}\Bigr)^{-1},
\label{P_Fock_ferm}\\
&P_{\rm ent}(E_\alpha)|_{\epsilon=-1} \!=\!
\Bigl(\!C_{2/f\!-\!1+[\sqrt{1\!-\!f\!+\!2f
E_\alpha/\hbar\omega_\alpha}\!-\!1]/(2f)}^{[\sqrt{1\!-\!f\!+\!2f
E_\alpha/\hbar\omega_\alpha}-1]/(2f)}\!\Bigr)^{-1},\label{P_Fock_bos}
\end{align}
the definition intervals being the same as for the entanglement
entropy, see~(\ref{S_1quas_ferm}) and (\ref{S_1quas_bos}).
 The functions~(\ref{P_Fock_ferm}) and (\ref{P_Fock_bos}) of energy
are presented graphically in~Fig.~\ref{fig11} and Fig.~\ref{fig12}
 correspondingly. Notice the peculiar shape of curves
 in~Fig.~\ref{fig11} (non-monotonic behavior, with two pieces of
 monotonicity for each curve).

\begin{figure}[h!]
\includegraphics[width = 0.8\columnwidth]{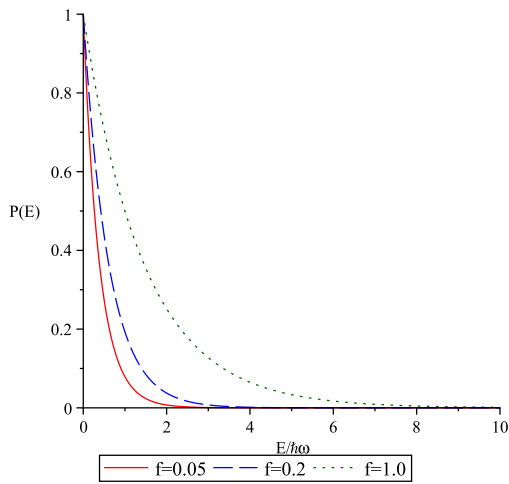}
\vskip-3mm\caption{\footnotesize Dependence of the purity $P$ on
energy $E$ for multi-quasibosonic system with one quasiboson per
mode: the case $\epsilon=\!+\!1$ of fermionic constituents.}
\label{fig13}
\end{figure}

\begin{figure}[h!]
\includegraphics[width = 0.8\columnwidth]{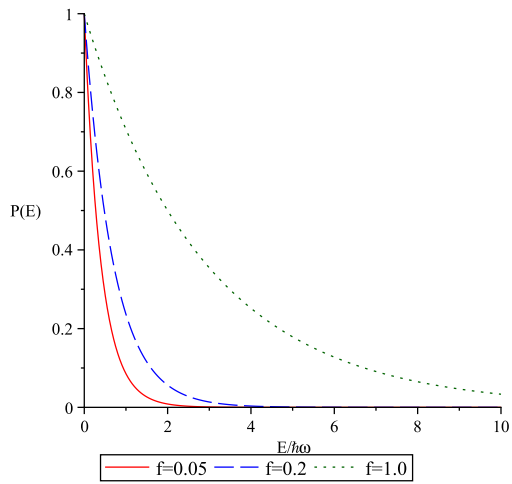}
\vskip-3mm\caption{\footnotesize Dependence of the purity $P$ on
energy $E_\alpha$ for multi-quasibosonic system with one quasiboson
per mode: the case $\epsilon=\!-\!1$ of bosonic constituents. }
\label{fig14}
\end{figure}

Likewise, for {\it multi-quasibosonic states with one quasiboson per
mode}, using the expression for purity calculated
in~\cite{GM_Entang} we easily find
\begin{equation}
P(E) = \exp(- S_{\rm ent}(E)).
\end{equation}
 The corresponding plots are now placed in~Figs.\ref{fig13}, \ref{fig14}.
As Figs.~\ref{fig12}-\ref{fig14} demonstrate, purity is falling from
its maximal possible value $P=1$ (zero entanglement) attained at
$E=\frac{\hbar\omega}{2}$ implying absence of quanta, to $P=0$ at
very large energies of many-particle states. A peculiar behavior of
purity as function of energy is seen in Fig.~\ref{fig11}: purity
drops from $P=1$ with energy growing to some maximum $E_{\rm
max}(f)$ (the latter is determined by the parameter $f$), then
further decreases from $P_f\equiv P(E_{\rm max}(f))$ with the energy
lowering from $E_{\rm max}(f)$ to smallest values. It is tempting to
interpret such existence of two regimes as follows: both addition
and subtraction~\cite{Kurzynski},\cite{Bartley} of quanta (of
quasiboson) can result in lowering purity. The two regimes, linked
with existence of two branches, differ in the starting value of
purity $P$ (whether it is zero or $P(E_{\rm max})$).

\section{Discussion}\label{sec:discussion}

In conclusion, let us make some comments on the obtained
dependencies of entanglement entropy on energy, and their
visualization with the corresponding plots. For the state of one
composite boson realized by deformed oscillator, using the
Hamiltonian~(\ref{Hamiltonian}) we find that the entanglement
entropy monotonically grows with energy if the components are
fermions, and decreases if the components are bosons
(Figures~\ref{fig1} and~\ref{fig2}).

We infer that for larger energies two-fermion quasiboson becomes
more tightly bound whereas two-boson quasiboson becomes less bound.
In the both cases the energy $E=\frac32 \hbar\omega$ corresponds to
the most entangled quasiboson which here shows itself as most close
to pure boson.

If we compare the case of a two-fermion quasiboson state with the
case of hydrogen atom viewed as a two-fermion composite almost
boson, we observe that in case of $H$-atom the dependence of the
entanglement entropy on energy shows decreasing and thus strongly
differs from the two-fermionic quasiboson case (compare
Fig.~\ref{fig1} and Fig.~\ref{fig3}). The reason may be rooted in
the specified proton-electron interaction and/or in the
non-elementary nature of proton, one of the two constituent fermions
of $H$-atom (see also last paragraphs in Sec. 3).

In the case of multi-quasiboson state for a single fixed mode and
when there are two fermionic components, we observe two branches --
one decreasing and the other increasing, see~Fig.~\ref{fig6}  and
the last but one paragraph of this section. For the rest of the
considered multi-quasibosonic states with fermionic components (the
fixed one mode case, or with one quasiboson per mode, or coherent
state) the entanglement entropy is monotonously growing with energy,
see Figs.~\ref{fig7}-\ref{fig10}, while purity is falling
monotonically as in Figs.~\ref{fig12}-\ref{fig14}, or with some
peculiarities (two regimes or branches of monotonicity,
see~Fig.~\ref{fig11} and the end of Sec. V).

What about the role of deformation parameter~$f$? We have quite
natural feature: the entanglement entropy is rising with decreasing
values of~$f$, i.e. with the  approaching to truly bosonic behavior,
either for the Fock states at fixed mode or for the coherent states.

For varying energy, from figures~\ref{fig7}-\ref{fig10}
 we find: the two-fermion quasiboson state, multi-quasibosonic
states for two-fermion quasibosons within a fixed mode,
multi-quasibosonic states with one quasiboson per mode, and the
coherent ones are the more entangled {\it the larger is the energy}.
This suggests a possibility to enhance the entanglement (i.e. its
entropy) by increasing energy of the (multi-)quasiboson state. The
states of two-boson quasibosons {\it show an opposite behavior} as
they are less entangled for larger energies. From the physics
viewpoint we thus have unambiguous relation between the degree
(strength) of entanglement and say the energy level of the
considered multi-quasibosonic states.


In some cases the dependencies obtained above, e.g. those in
Fig.~\ref{fig1} and Figs.~\ref{fig7}-\ref{fig10}, can be viewed in
the context of {\it entanglement production or enhancement},
(see~\cite{Weder}, \cite{Leon}) and this provides another possible
physical implication of our results. As seen, the entanglement
becomes greater with increasing energy (particle addition?) for the
listed cases. Unlike those ones, in the case of two-boson
quasibosons the {\it entanglement creation} is observed when the
energy is decreased. That is, when the energy of system is lowered
(particle subtraction?), the entanglement entropy grows. Possibly,
this could be checked for some physical examples of the treated
systems, especially from the particle addition/subtraction
viewpoint~\cite{Kurzynski},\cite{Bartley}.

Let us make few more remarks on possible experimental verification
of the obtained results. That may concern the dependencies shown
e.g. in Fig.~\ref{fig1} and Fig.~\ref{fig2}. As for the first case
(two-fermionic quasiboson) one may consider electron-electron or
electron-hole composites (excitons). To test the properties for the
two-boson composite we could take bi-photons or the H-molecule $H_2$
for the corresponding relevant experiments. Besides,
multi-quasibosonic dependencies presented in Fig.~\ref{fig8} and
Fig.~\ref{fig9} may also be of practical or physical interest.

At last, let us note the intriguing appearance of ``bifurcations''
(or existence of two branches) that are in Fig.~\ref{fig6} and
Fig.~\ref{fig11}. Which of the branches is physically realized could
be an intriguing possible issue for verification. Say,
$2k$-electron, $2k$-photon, $k$-exciton systems, etc. could be used
for dedicated experiments.
 Besides the quasibosonic states studied in this work,
non-pure e.g. thermal states of quasibosons are also of interest for
future analysis.

We hope to extend the above treatment, the obtained results, and
physical implications, to more complex quasi-boson (or
quasi-fermion) systems in our forthcoming works, and also to compare
with another real physical examples (like the $H$-atom considered
above).
\par
{\bf Acknowledgments}. We are grateful to I.V. Simenog and N.Z.
Iorgov for valuable discussion. Thanks are also to the referees
whose constructive remarks resulted in improved presentation.
 The research was partially supported by the Special Program of the Division of Physics and
Astronomy of NAS of Ukraine.

\appendix
\renewcommand\thesection{\Alph{section}}

\section{Derivation of the entanglement entropy for Hydrogen atom}

Transforming the sum  in~(\ref{S_Hydr}) into integral (that implies
very large volume $V$)  and substituting the Hydrogen wavefunctions
from~(\ref{phi_nlm}), for~$S_{\rm ent}$ we obtain
\begin{multline}
S_{\rm ent} \!=\! -\! \int\! \frac{Vd^3{\bf p}}{(2\pi\hbar)^3} \frac{(2\pi\hbar)^3}{V} |\phi_{{\bf p}nlm}|^2 \ln\Bigl(\frac{(2\pi\hbar)^3}{V} |\phi_{{\bf p}nlm}|^2\Bigr) =\\
\!=\! -\! \int\! \sin\theta_p d\theta_p p^2 dp \frac{(2l\!+\!1)(l\!-\!m)!}{2(l\!+\!m)!} |P^m_l(\cos \theta_p)|^2\cdot\\
\cdot\frac{(\pi 2^{2l+4}l!)^2}{(\gamma h)^3}\, \frac{n(n\!-\!l\!-\!1)!}{(n\!+\!l)!} \frac{\xi^{2l}}{(\xi^{2}\!+\!1)^{2l+4}} \Bigl(C^{l+1}_{n-l-1}\Bigl(\frac{\xi^2\!-\!1}{\xi^2\!+\!1}\Bigr)\Bigr)^2\cdot\\
\cdot \ln\biggl(\frac{(2\pi\hbar)^3}{V} \frac{1}{2\pi} \frac{(2l\!+\!1)(l\!-\!m)!}{2(l\!+\!m)!} |P^m_l(\cos \theta_p)|^2 \, \frac{(\pi 2^{2l+4}l!)^2}{(\gamma h)^3}\cdot\\
\cdot \frac{n(n\!-\!l\!-\!1)!}{(n\!+\!l)!}
\frac{\xi^{2l}}{(\xi^{2}\!+\!1)^{2l+4}}
\Bigl(C^{l+1}_{n-l-1}\Bigl(\frac{\xi^2\!-\!1}{\xi^2\!+\!1}\Bigr)\Bigr)^2\biggr)
\mathop{=}\limits^{t=\cos\theta_p}\\
\mathop{=}\limits^{t=\cos\theta_p} - \int\limits_{-1}^{1} dt |P^m_l(t)|^2\int\limits_{0}^{\infty} \frac{\xi^{2l+2}d\xi}{(\xi^{2}\!+\!1)^{2l+4}} \, \frac{(2l\!+\!1)(l\!-\!m)!}{(l\!+\!m)!}\cdot\\
\cdot \frac{2^{4l+4}(l!)^2}{\pi}
\frac{n(n\!-\!l\!-\!1)!}{(n\!+\!l)!}\,
\Bigl(C^{l+1}_{n-l-1}\Bigl(\frac{\xi^2\!-\!1}{\xi^2\!+\!1}\Bigr)\Bigr)^2\cdot\\
\cdot \ln\biggl(|P^m_l(t)|^2
\frac{(2l\!+\!1)(l\!-\!m)!}{(l\!+\!m)!}\frac{\pi
2^{4l+6}(l!)^2}{V(1/na_0)^3}
\frac{n(n\!-\!l\!-\!1)!}{(n\!+\!l)!}\cdot\\
\cdot\frac{\xi^{2l}}{(\xi^{2}\!+\!1)^{2l+4}}
\Bigl(C^{l+1}_{n-l-1}\Bigl(\frac{\xi^2\!-\!1}{\xi^2\!+\!1}\Bigr)\Bigr)^2\biggr),
\end{multline}
Recall that $C^{l+1}_{n-l-1}(...)$ is the Gegenbauer polynomial. For
convenience, introduce new variable
$x=\frac{\xi^2\!-\!1}{\xi^2\!+\!1}$ and the function $G_{nl}(x) =
(1-x^2)^l (1-x)^4 \bigl(C^{l+1}_{n-l-1}(x)\bigr)^2$. Then we arrive
at the expression:
\begin{multline}
S_{\rm ent} \!=\!  -\! \frac{(2l\!+\!1)(l\!-\!m)!}{(l\!+\!m)!} \frac{2^{2l}(l!)^2}{\pi} \frac{n(n\!-\!l\!-\!1)!}{(n\!+\!l)!} \int\limits_{-1}^{1} dt |P^m_l(t)|^2 \cdot\\
\int\limits_{-1}^{1}\! dx \frac{\sqrt{1\!-\!x^2}}{(1\!-\!x)^3} G_{nl}(x) \ln\biggl(\!|P^m_l(t)|^2\frac{(2l\!+\!1)(l\!-\!m)!}{(l\!+\!m)!}\frac{\pi 4^{l+\!1}(l!)^2}{V(na_0)^{-\!3}}\\ \frac{n(n\!-\!l\!-\!1)!}{(n\!+\!l)!} G_{nl}(x)\!\biggr)
\!=\! - \frac{(2l\!+\!1)(l\!-\!m)!}{(l\!+\!m)!} \frac{4^l(l!)^2}{\pi} \frac{n(n\!-\!l\!-\!1)!}{(n\!+\!l)!}\cdot\\
\int\limits_{-1}^{1} dt |P^m_l(t)|^2\int\limits_{-1}^{1} dx \frac{\sqrt{1\!-\!x^2}}{(1\!-\!x)^3} G_{nl}(x)
\cdot\biggl\{\ln\Bigl(\frac{(2l\!+\!1)(l\!-\!m)!}{(l\!+\!m)!}\cdot\\
\frac{\pi 4^{l+\!1}(l!)^2}{V(na_0)^{-\!3}} \frac{n(n\!-\!l\!-\!1)!}{(n\!+\!l)!}\Bigr) \!+\! \ln |P^m_l(t)|^2 \!+\! \ln G_{nl}(x)\biggr\}.
\end{multline}
Using normalization condition for $P^m_l(t)$, that is $\int_{-1}^{1}
\bigl(P^m_l(t)\bigr)^2 dt = \frac{2(l+m)!}{(2l+1)(l-m)!}$, for
$S_{\rm ent}$ we derive:
\begin{multline}\label{S_ent1}
S_{\rm ent} = - \frac{2^{2l\!+1}(l!)^2}{\pi} \frac{n(n\!\!-\!\!l\!\!-\!\!1)!}{(n\!+\!l)!}  \ln\Bigl[\frac{(2l\!\!+\!\!1)(l\!\!-\!\!m)!}{(l\!+\!m)!}\frac{4\pi 4^l(l!)^2}{V(na_0)^{-3}}\cdot\\
\cdot\frac{n(n\!\!-\!\!l\!\!-\!\!1)!}{(n\!+\!l)!}\Bigr] \!\int\limits_{-1}^{1}\! dx \frac{\sqrt{1\!-\!x^2}}{(1\!-\!x)^3} G_{nl}(x) - \frac{(2l\!+\!1)(l\!-\!m)!}{(l\!+\!m)!} \frac{4^l(l!)^2}{\pi} \cdot\\
\frac{n(n\!-\!l\!-\!1)!}{(n\!+\!l)!} \!\int\limits_{-1}^{1}\! dx \frac{\sqrt{1\!-\!x^2}}{(1\!-\!x)^3} G_{nl}(x) \!\int\limits_{-1}^{1}\! dt |P^m_l(t)|^2 \ln |P^m_l(t)|^2 -\\
\!-\! \frac{4^l(l!)^2\!}{\pi/2} \frac{n(n\!-\!l\!-\!1)!}{(n\!+\!l)!}
\!\int\limits_{-1}^{1}\! dx \frac{\sqrt{1\!-\!x^2}\!}{(1\!-\!x)^3}
G_{nl}(x) \ln G_{nl}\!(x).
\end{multline}
Using orthonormalization condition and recurrence relation for Gegenbauer polynomials, we have
\begin{equation}
\int\limits_{-1}^{1} dx \frac{\sqrt{1\!-\!x^2}}{(1\!-\!x)^3} G_{nl}(x) \!=\! \frac{\pi 2^{-1-2l} (n+l)!}{(n\!-\!l\!-\!1)!n (l!)^2}.
\end{equation}
Then from~(\ref{S_ent1}) we finally obtain
\begin{multline}
S_{\rm ent} = - \ln\Bigl[\frac{(2l+1)(l-m)!}{(l+m)!}\frac{4\pi 2^{2l}(l!)^2}{V(na_0)^{-3}} \frac{n(n-l-1)!}{(n+l)!}\Bigr]-\\
- \frac{(2l+1)(l-m)!}{2(l+m)!} \, \int\limits_{-1}^{1} dt |P^m_l(t)|^2 \ln |P^m_l(t)|^2 -\\
\!-\!\frac{4^l(l!)^2}{\pi/2} \frac{n(n\!-\!l\!-\!1)!}{(n+l)!}
\!\int\limits_{-1}^{1}\! dx \frac{\sqrt{1\!-\!x^2}\!}{(1\!-\!x)^3}
G_{nl}\!(x) \ln G_{nl}\!(x).
\end{multline}

\bibliographystyle{apsrev4-1}
\bibliography{references}

\end{document}